\documentclass[reprint, amsmath,amssymb, aps,superscriptaddress]{revtex4-2}
\usepackage{graphicx}% Include figure files
\usepackage{dcolumn}% Align table columns on decimal point
\usepackage{bm}% bold math
\usepackage[english]{babel} % languages, maybe more than one
\usepackage{lmodern} % use latin modern fonts 
\usepackage[dvipsnames]{xcolor}
\usepackage{hyperref} % turn all your internal references into hyperlinks
\usepackage{natbib} % Journal style bibliography managmenet. Bibliography files in the bibtex syntax are expected

\usepackage{amsmath, amsthm, amscd, amssymb} % the AMS packages:
\usepackage{braket}
\usepackage[T1]{fontenc} % accents and so on
\usepackage{geometry} % for document formatting
\usepackage{lineno}
\makeindex
\begin{document}

\title{{\bf  Strong Optomechanical  Coupling at Room Temperature by Coherent Scattering}}

\author{Andr\'es de los R\'ios Sommer}
\affiliation{ICFO-Institut de Ciencies Fotoniques, The Barcelona Institute of Science and Technology, 08860 Castelldefels (Barcelona), Spain}

\author{Nadine Meyer}\thanks{Corresponding author: nmeyer@ethz.ch}%
\affiliation{ICFO-Institut de Ciencies Fotoniques, The Barcelona Institute of Science and Technology, 08860 Castelldefels (Barcelona), Spain}

\author{Romain Quidant}
\affiliation{ICFO Institut de Ciencies Fotoniques, Mediterranean Technology Park, 08860 Castelldefels (Barcelona), Spain}
\affiliation{ICREA-Instituci\'{o} Catalana de Recerca i Estudis Avan\c{c}ats, 08010 Barcelona, Spain}
\affiliation{Nanophotonic Systems Laboratory, Department of Mechanical and Process Engineering, ETH Zurich, 8092 Zurich, Switzerland}

\begin{abstract}
\noindent \textbf{Abstract} Quantum control of a system requires the manipulation of quantum states faster than any decoherence rate. For mesoscopic systems, this has so far only been reached by few cryogenic systems. An important milestone towards quantum control is the so-called strong coupling regime, which in cavity optomechanics corresponds to an optomechanical coupling strength larger than cavity decay rate and mechanical damping. Here, we demonstrate the strong coupling regime at room temperature between a levitated silica particle and a high finesse optical cavity. Normal mode splitting is achieved by employing coherent scattering, instead of directly driving the cavity.   The coupling strength achieved here approaches three times the cavity linewidth, crossing deep into the strong coupling regime. Entering the strong coupling regime is an essential step towards quantum control with mesoscopic objects at room temperature. 
\end{abstract}

%%%%%%%%%%%%%%%%%%%%%%%%%%%%%%%%%%%%%%%%%%%%%%%%%%%%%%%%%%%%%%%%%%%%%%%%%%%%%%%%%%%%%%%%%%%%%%%%%%%

\maketitle

%%%%%%%%%%%%%%%%%%%%%%%%%%%%%%%%%%%%%%%%%%%%%%%%%%%%%%%%%%%%%%%%%%%%%%%%%%%%%%%%%%%%%%%%%%%%%%%%%%%

%\paragraph*{Introduction}
\subsection*{Introduction} 
\noindent Laser cooling has revolutionised our understanding of atoms, ions and molecules. 
%Lately, after a decade of experimental and theoretical efforts\cite{Chang2010, Romero-Isart2010,Kiesel2013,Asenbaum2013,Millen2015,Fonseca2016,Meyer2019,Delic2019}, laser cooling has also enabled the motional ground state cooling of levitated Silica nanoparticles \cite{Delic2020}. 
Lately, after a decade of experimental and theoretical efforts employing the same techniques \cite{Chang2010, Romero-Isart2010,Kiesel2013,Asenbaum2013,Millen2015,Fonseca2016,Meyer2019,Delic2019}, the motional ground state of levitated silica nanoparticles at room temperature has been reported \cite{Delic2020}. 
While this represents an important milestone towards the creation of mesoscopic quantum objects, coherent quantum control of levitated nanoparticles \cite{Romero-Isart2011,Bateman2014a} %, as needed for quantum information processing, 
still remains elusive.\\
\noindent Levitated particles stand out among the plethora of optomechanical systems \cite{Aspelmeyer2014} due to their detachment, and therefore high degree of isolation from the environment. Their centre of mass, rotational and vibrational degrees of freedom \cite{Millen2020} %, rendering Q-factors of $10^8$ \cite{Gieseler2013}. 
make them attractive tools for inertial sensing \cite{Hempston2017}, rotational dynamics \cite{Arita2013,Kuhn2017,Monteiro2018,Reimann2018}, free fall experiments \cite{Hebestreit2018a}, exploration of dynamic potentials \cite{Rondin2017}, and are envisioned for testing macroscopic quantum phenomena at room temperature \cite{Marshall2003, Kleckner2008, Romero-Isart2010,Romero-Isart2011}.\\  
\noindent Recently, the centre-of-mass motion of a levitated particle has successfully been 3D cooled employing coherent scattering (CS) \cite{Windey2019,Delic2019}.  Cooling with CS is less sensitive to phase noise heating than actively driving the cavity \cite{Meyer2019,Delic2020a}, because optimal coupling takes place at the intensity node. Lately, this has enabled %originally suggested for atoms in an optical cavity \cite{Vuletic2001}, and finally
phonon occupation numbers of less than one %motional ground state cooling %of a levitated oscillator
\cite{Delic2020}. %CS has initially been suggested for atoms in an optical cavity \cite{Vuletic2001} and can be used to alter the spontaneous emission of photons.
\\
\noindent For controlled quantum experiments, such as the preparation of non-classical, squeezed \cite{Safavi2013,Riedinger2016} or entangled states \cite{Riedinger2018,Chen2020}, the particle's motional state needs to be manipulated faster than the absorption of a single phonon from the environment. A valuable but less stringent condition is the so-called strong coupling regime (SCR), where the optomechanical coupling strength $g$ between the mechanical motion of a particle and an external optical cavity exceeds the particle's mechanical damping $\Gamma_\text{m}$ and the cavity linewidth $\kappa$ ($g\gg \Gamma_\text{m}, \kappa$). The SCR presents one of the first stepping stones towards full quantum control and has been demonstrated in opto- and electromechanical systems \cite{Groblacher2009,Teufel2011a, Teufel2011}, followed by quantum-coherent control \cite{Verhagen2012c}.\\

%%%%%%%%%%%%%%%%%%%%%%%%%%%%%%%%%%%%%%%%%%%%%%%%%%%%%%%%%%%%%%%%%%%%%%%%%%%%%%%%%%%%%%%%%%%%%%%%%%%
%  
\noindent Here, we observe normal mode splitting (NMS) in SCR with levitated nanoparticles \cite{Dobrindt2008b}, as originally reported in atoms \cite{Thompson1992}. In contrast to previous experiments, we employ CS \cite{Vuletic2001,Gonzalez-Ballestero2019,Windey2019,Delic2019}. 
Our table top experiment offers numerous ways to tune the optomechanical coupling strength %acting on very isolated mechanical modes 
at room temperature, a working regime that is otherwise nearly exclusive to plasmonic nanocavities \cite{Chikkaraddy2016,Kleemann2017a}.\\

\subsection*{Results}
\begin{figure}
	\begin{center}
		\includegraphics[width=0.45\textwidth]{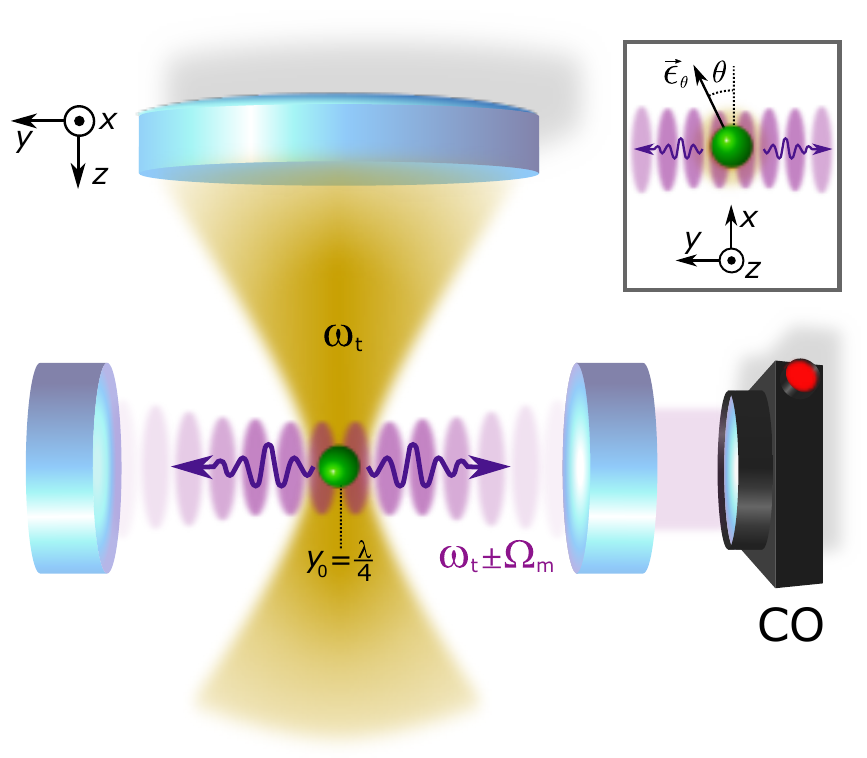}
		\caption{\footnotesize \textbf{Experimental setup:} An optical tweezers trap (yellow) levitates a silica nanoparticle inside a high finesse cavity. The trapping field is locked relatively to the cavity resonance $\omega_{\text{cav}}$ using Pound-Drever-Hall locking with a detuning $\Delta =\omega_{\text{t}} - \omega_{\text{cav}}$. A 3D piezo stage positions the particle precisely inside the cavity at variable $y_0$. The inset displays the linear trap polarisation axis along $\epsilon_\theta=\epsilon_\text{x}\cos{\theta} $. The rate of coherently scattered photons into the cavity mode (purple) depends on $y_0$, $\epsilon_\theta$, and $\Delta$. The transmitted cavity output field is monitored on a photodiode (CO) and the forward scattered trapping light is used to detect the particle motion (see Methods \ref{app:setup}).}
		\label{fig:01_setup}
	\end{center}
\end{figure}

\noindent \textbf{Experimental Setup for Levitation}
Our experimental setup is displayed in Fig.\ref{fig:01_setup}. A silica nanoparticle (green) of radius $R \approx 90$nm, mass $m = 6.4 \times 10^{-18}$kg and refractive index $n_\text{r}  = 1.45$ is placed in a cavity (purple) by an optical tweezers trap (yellow) with wavelength $\lambda_\text{t} = 2\pi/k_\text{t} = 1064~{\rm nm}$, power $P_\text{t}\simeq 150~{\rm mW}$, numerical aperture $\text{NA}=0.8$, and optical axis ($z$) perpendicular to the cavity axis ($y$). The trap is linearly polarised along the axis  defined as $\epsilon_{\theta} =\epsilon_{\text{x} }\cos{\theta} $ (see inset in Fig.\ref{fig:01_setup}). \\ 
The nanoparticle's eigenfrequencies $\Omega_{\text{x,y,z}} = 2\pi\times ({\rm 172kHz,\: 197kHz,\: 56kHz})$ are non-degenerate due to tight focusing. The trap is mounted on a nano-positioning stage allowing for precise 3D placement of the particle inside the low loss, high finesse Fabry-P\'{e}rot cavity with a cavity linewidth $\kappa \approx 2\pi\times 10$kHz, cavity finesse $F = 5.4\times 10^5$ and free spectral range $\Delta \omega_{\text{FSR}} =\pi c /L_{\text{c}} = 2\pi\times 5.4$GHz. The relative detuning $\Delta = \omega_\text{t} - \omega_{\text{c}}$ between the trap and the cavity resonance is tunable. The intracavity photon number $n_{\text{cav}}$ is estimated from the transmitted cavity power $P_{\text{out}}$ (CO in Fig.\ref{fig:01_setup}), and the particle position displacement is measured by interfering the scattered light with a co-propagating reference beam \cite{Gieseler2014a}. % balanced detection of the forward-scattered light (for more details see Methods \ref{app:setup}). \\ 
In CS, scattering events from the detuned trapping field, locked  at $\Delta$, populate the cavity. This contrasts the approach of actively driving the cavity \cite{Kiesel2013,Meyer2019,Delic2020a}. 
A particle in free space, solely interacting with the trapping light, Raman scatters photons into free space and the energy difference between incident and emitted light equals %units of motional quanta 
$\pm \hbar \Omega_\text{m}$ with $\text{m={x,y,z}}$. %,according to energy and momentum conservation. 
In this case photon up and down conversion are equally probable \cite{Tebbenjohanns2020}.
%In absence of the cavity, the particle scatters photons into free space and the energy difference between incident and emitted light equals multiple units of motional quanta, according to energy and momentum conservation. 
The presence of an optical cavity alters the density of states of electromagnetic modes and enhances the CS into the cavity modes through the Purcell effect. % Each scattering event transfers units of the photon recoil energy $E_{rec} = \hbar^2 k^2/(2m)$ to the particle.  \\
If trap photons are red (blue) detuned with respect to the cavity resonance, the cavity enhances photon up (down) conversion and net cooling (heating) takes place.\\

%%%%%%%%%%%%%%%%%%%%%%%%%%%%%%%%%%%%%%%%%%%%%%%%%%%%%%%%%%%%%%%%%%%%%%%%%%%%%%%%%%%%%%%%%%%%%%%%%%%%
%
%%%%%%%%%%%%%%%%%%%%%%%%%%%%%%%%%%%%%%%%%%%%%%%%%%%%%%%%%%%%%%%%%%%%%%%%%%%%%%%%%%%%%%%%%%%%%%%%%%%%%%%%%%%

\noindent \textbf{Coherent Scattering Theory}
%\paragraph*{Coherent Scattering Theory}
\noindent In order to estimate the corresponding optomechanical coupling strength in CS, we follow %the theory for CS in Levitodynamics derived in  
\cite{Gonzalez-Ballestero2019}. The interaction Hamiltonian %$\hat{H}_{int}$ 
for a polarizable particle interacting with an electric field $\rm \pmb{\rm E}(\pmb{\rm R})$ is given by
%\begin{equation}
$\hat{\rm H}_{\text{int}} = - \frac{1}{2} \alpha \pmb{\rm E}^2(\pmb{\rm R}) $
%\end{equation}
with the particle polarizability $\alpha = 4\pi \epsilon_0 R^3 \frac{n_\text{r}^2-1}{n_\text{r}^2 +2}$ and vacuum permittivity $\epsilon_0$. %Here we only consider low absorptive particles smaller than the optical wavelength $\lambda_t \gg R $. %The interaction Hamiltonian needs to be evaluated at the particle position $\pmb{\hat{R}}$. 
%In the case of CS t
The total electric field consists of the trap ($\pmb{\rm E}_{\text{tr}}(\pmb{\rm R})$), cavity ($\pmb{\rm E}_{\text{cav}}(\pmb{\rm R})$) and free space electromagnetic modes ($\pmb{\rm E}_{\text{free}}(\pmb{\rm R})$) yielding the interaction Hamiltonian %The trap and the intracavity field are assumed as hermite gaussian beams, where the coherent state of the trap is always much higher populated than the intracavity field ($n_{trap} \gg n_{\text{cav}}$).  

\begin{eqnarray}
\hat{\rm H}_{\text{int}} &=& - \frac{1}{2} \alpha \left[\pmb{\rm E}_{\text{tr}}(\pmb{\rm R}) + \pmb{\rm E}_{\text{cav}}(\pmb{\rm R})  + \pmb{\rm E}_{\text{free}}(\pmb{\rm R})  \right] ^2 \label{eq:Hint1}\\
   &\approx& \hat{\rm H}_{\text{CS}} + \hat{\rm H}_{\text{DR}} + \hat{\rm H}_{\text{CAV}} \label{eq:Hint2}
\end{eqnarray}

\noindent where $\pmb{\rm E}_{\text{cav}}(\pmb{\rm R})$  and $\pmb{\rm E}_{\text{free}}(\pmb{\rm R})$ are only populated by scattering events from the particle ($n_{\text{trap}} \gg n_{\text{cav}}$ with $n_{\text{trap}}$ ($n_{\text{cav}}$) being the number of trap (cavity) photons). As can be seen from Eq. \ref{eq:Hint1}, the interaction Hamiltonian consists of six terms of which only the two terms proportional to $\pmb{\rm E}_{\text{tr}}(\pmb{\rm R})\pmb{\rm E}_{\text{cav}}(\pmb{\rm R})$ and $\pmb{\rm E}_{\text{cav}}(\pmb{\rm R})^2$ are relevant for the following discussion \cite{Gonzalez-Ballestero2019}. The former one gives rise to the optomechanical coupling by CS, and the latter to the coupling achieved by actively driving the cavity. %, if the cavity is highly populated ($n_{cav} \gg 0$). 
The term $\propto \pmb{\rm E}_{\text{tr}}^2(\pmb{\rm R})$ gives rise to the trapping potential, % and allows us to quantise the motion as $(n+1/2)\:\Omega_{x,y,z}$, 
while the term $\propto\pmb{\rm E}_{\text{tr}}(\pmb{\rm R})   \pmb{\rm E}_{\text{free}}(\pmb{\rm R})$ causes recoil heating \cite{Jain2016,Gonzalez-Ballestero2019}, which can be neglected for the moderate vacuum conditions presented here \cite{Jain2016,Meyer2019}. % Although both terms cannot be neglected as such, they are of no importance for the following discussion. 
The remaining two terms can be safely neglected according to \cite{Gonzalez-Ballestero2019}. \\
In the following, we use the simplified interaction Hamiltonian given by Eq.\ref{eq:Hint2} where we separate the parts contributing to the optomechanical coupling due to CS $\hat{\rm H}_{\text{CS}}$, active driving $\hat{\rm H}_{\text{DR}}$, and population of the intracavity field $\hat{\rm H}_{\text{CAV}}$ (see Methods \ref{app:EqMotionPSD}).
%In general, CS allows for 3D cooling \cite{Windey2019,Delic2019,Gonzalez-Ballestero2019} with the advantage to be less sensitive to phase noise heating than radiation pressure cooling \cite{Meyer2019,Delic2020a}, since optimal cooling takes place at the intensity node. 
%Additionally we safely neglect photon recoil as decoherence source since our experiments take place in moderate vacuum \cite{Jain2016,Meyer2019}.
\\
For the measurements presented here, the trap is $x$-polarised with $\theta = 0$ %, perpendicular to the cavity and trap axis
(see inset Fig.\ref{fig:01_setup}). %Thus our interaction Hamiltonian takes the form Eq.\ref{eq:Hint2} where we separated the contributions stemming from optomechanical coupling due CS $\hat{H}_{\text{CS}}$, radiation pressure $\hat{H}_{\text{RP}}$, and population of the intracavity field $\hat{H}_{\text{CAV}}$ (see Methods \ref{app:EqMotionPSD}). 
This  simplifies $\hat{\rm H}_{\text{CS}}$ to 
$ \hat{\rm H}_{\text{CS}}= - \hbar [g_{\text{y}} (\hat{a}^\dagger +\hat{a})  (\hat{b_y}^\dagger + \hat{b_y}) + g_{\text{z}} (\hat{a}^\dagger - \hat{a}) (\hat{b_z}^\dagger + \hat{b_z})] $
%\begin{eqnarray}
%\frac{\hat{H}_{\text{CS}}}{\hbar}&=& - g_{y} (\hat{a}^\dagger +\hat{a})  (\hat{b_y}^\dagger + \hat{b_y}) -  g_{z} (\hat{a}^\dagger - \hat{a}) (\hat{b_z}^\dagger + \hat{b_z})  \nonumber \\  
%\frac{\hat{H}_{\text{RP}}}{\hbar}&=& - g^\text{rp}_y\:\hat{a}^\dagger \hat{a}\:  (\hat{b_y}^\dagger + \hat{b_y})  \nonumber \\
%\frac{\hat{H}_{\text{CAV}}}{\hbar} &=& -  \frac{G_{\perp}}{2} (\:\hat{a}^\dagger + \hat{a}) \cos{\phi}\nonumber
%\end{eqnarray}
where $\hat{a}$ ($\hat{a}^\dagger$) is the photon annihilation (creation) operator and  $\hat{b}$ ($\hat{b}^\dagger$) is the phonon annihilation (creation) operator. 
\noindent The CS optomechanical coupling strengths $g_{\text{y,z}}$ are
\begin{equation}
\begin{bmatrix} \label{eq:gy}
g_y\\
g_z
\end{bmatrix}
=  \frac{1}{2}\begin{bmatrix}
&G_{\perp} \:k_\text{c} \: y_{\text{zpf}} \:\sin{\phi}\\
-i& G_{\perp} \: k_\text{t} \: z_{\text{zpf}} \:\cos{\phi}\\
\end{bmatrix}
\end{equation}
with cavity wavevector $k_\text{c} = 2\pi/\lambda_\text{c}$, zero-point fluctuations $y_{\text{zpf}},z_{\text{zpf}} =  \sqrt{\frac{\hbar}{2\: m\: \Omega_{\text{y,z}}}}$ and $\phi = 2\pi y_0/\lambda_\text{c}$, with $y_0$ being the particle position along the cavity axis and $y_0 = \lambda_\text{c}/4$ corresponding to the intensity minimum.\\ \noindent The optical cavity resonance frequency shift caused by a particle located at maximum intensity of the intracavity standing wave is  
%\begin{equation}\label{eq:G}
$G_{\perp} = \alpha E_0 \sqrt{\frac{\omega_\text{c}}{2\hbar \epsilon_0 V_\text{c}}} $
%\end{equation}
with cavity mode volume $V_\text{c} = \pi \text{w}_\text{c}^2 L_\text{c}/4$, cavity waist $\text{w}_\text{c}$, cavity length $L_\text{c}$, and $\omega_\text{c} = 2\pi c/\lambda_\text{c}$. The trap electric field is $E_0 = \sqrt{\frac{4 P_\text{t}}{\pi \epsilon_0 c \text{w}_\text{x} \text{w}_\text{y}}}$ with trap waists $\text{w}_\text{x}$ and $\text{w}_\text{y}$.\\
Due to the intracavity standing wave, the optomechanical coupling strength has a sinusoidal dependence on $y_0$ with opposite phase for $g_\text{y}$ and $g_\text{z}$. %as 
%\begin{equation} \label{eq:gy_sin}
%\lvert g_{y} \rvert     = \lvert g_{y \max} \sin(2\pi y_0/\lambda_c)\rvert
%\end{equation} 
%where 
In contrast, $g_\text{x} = 0$ if $\theta = 0 $.%The optomechanical coupling strength  perpendicular to the cavity and trap axis is $g_x= 0$ for $\theta = 0 $.
\\
For clarity, we limit the discussion to coupling along the cavity axis (y), such that $\Omega_\text{m} = \Omega_\text{y}$ and $g = g_\text{y}$. %, despite that our setup allows for high control of all the aforementioned parameters.
 Similar results can be obtained for the other directions $x,z$ with the same level of control.\\ 
The maximum expected coupling strength from CS is $g_{\text{y}}^{\text{max}} = G_{\perp} k_{\text{c}} \:y_{\text{zpf}} = 2\pi\times 31.7\text{kHz}$ for our parameters. However, we displace the particle by  $\delta z \approx 40\mu\text{m}$ from the cavity centre for better experimental stability. Hence, our expected optomechanical coupling strength is reduced by $\approx 30\%$ down to $g_{\text{y}}^{\text{th}} = 2\pi\times 22.6\text{kHz}$, enabling the SCR with $g_\text{y} > \kappa$. Despite the fact that this value is a factor of $\approx 3$ lower than previously reported \cite{Delic2020}, the deep SRC with  $g > \kappa$ remains unaccomplished. \\

%
%%%%%%%%%%%%%%%%%%%%%%%%%%%%%%%%%%%%%%%%%%%%%%%%%%%%%%%%%%%%%%%%%%%%%%%%%%%%%%%%%%%%%%%%%%%%%%%%%%%%%
\begin{figure}
	\begin{center}
		\includegraphics[width=0.45\textwidth]{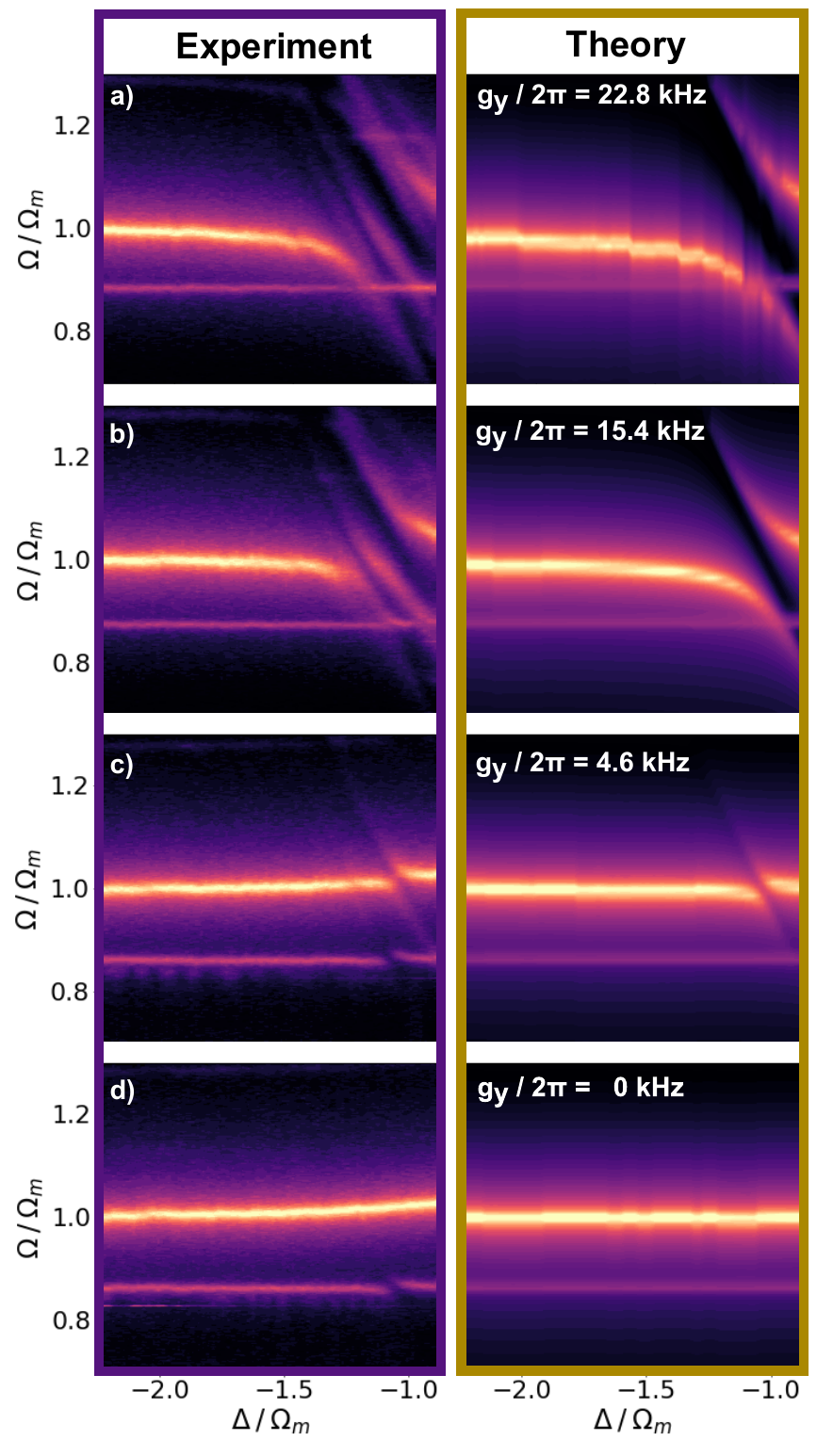}
		\caption{\footnotesize \textbf{Normal Mode Splitting:} Particle's position power spectral density PSD$(\Omega)$ versus $\Delta$ for different $y_0$ and therefore various $g_\text{y}$. Experimental data is displayed on the left, and theory on the right. The bare mechanical (optical) modes correspond to horizontal (diagonal) lines. Maximum normal mode splitting of $2g_\text{y}$ is observed at $\Delta = - \Omega_\text{m}$ yielding a value of  (\textbf{a}) $g_\text{y} =2\pi\times 22.8\text{kHz} = 2.3\kappa$, where $y_0\approx \lambda_\text{c}/4$ is close to the intensity minimum %at $\phi = 0.54\pi$ 
		(see Eq.\ref{eq:gy}). (\textbf{b}) When the particle is moved by $\delta y_0 \approx 0.12\lambda_\text{c}$%\approx 125$nm, corresponding to $\phi \approx 0.3\pi$
		, the coupling reduces to $g_\text{y} = 2\pi\times  15.4 \text{kHz} = 1.5\kappa$. (\textbf{c}) Normal mode splitting is still visible at $\delta y_0\approx 0.2\lambda_c$%205$nm, corresponding to $\phi \approx 0.15\pi$
		, yielding $g_\text{y} =2\pi\times 4.6\text{kHz} = 0.46\kappa$. %By varying $y_0$ by $\delta y_0\approx 125$nm, corresponding to $\phi \approx 0.3\pi$, we reduce $g_y$ to $g_y = 2\pi\times  15.4 \text{kHz} = 1.5\kappa$. (\textbf{c}) NMS is still expected at $g_y \approx \kappa/4$ confirmed by $g_y =2\pi\times 4.6\text{kHz} = 0.46\kappa$ with $\delta y0\approx 205$nm, corresponding to $\phi \approx 0.15\pi$.  
		(\textbf{d}) At the intensity maximum, corresponding to a shift of $\delta y_0 %\approx 287\text{nm} 
		\approx \lambda_\text{c}/4$ and $g_\text{y} =  0\text{kHz}$, the normal mode splitting vanishes and we only see a shift of $\delta \Omega_\text{m}\approx 2\pi\times$5kHz in the mechanical frequency due to the increased intracavity photon number (see Supplementary Fig.1%\ref{fig:cavintvsx2020-03-04}
		). In general, we observe a good agreement between experimental data and theory. We attribute discrepancies to a second cross polarised cavity mode inducing a second normal mode splitting (for more details see SI%\ref{app:MultiModeCoupl}
		).}
		\label{fig:02}
	\end{center}
\end{figure}
%%%%%%%%%%%%%%%%%%%%%%%%%%%%%%%%%%%%%%%%%%%%%%%%%%%%%%%%%%%%%%%%%%%%%%%%%%%%%%%%%%%%%%%%%%%%%%%%%%%%%%
%

\noindent \textbf{Transition to the Strong coupling regime} In the weak coupling regime $g<\kappa$, the Lorentzian shaped spectra of our mechanical oscillator displays a single peak at its resonance frequency $\Omega_\text{m}$.   When $g$ increases, the energy exchange rate between optical and mechanical mode grows until the SCR is reached at $g>\kappa/4$ \cite{Dobrindt2008b}. In the SCR, the optical and mechanical mode hybridise, which gives rise to two new eigenmodes at shifted eigenfrequencies $\Omega_{\pm}$ (see Eq.\ref{eq:Sxx}).  At this point the energy exchange in between the optical and mechanical mode is faster than the decoherence rate of each individual mode.
% While the weak coupling regime is best for cooling %with radiation pressure
%, the SCR is needed to perform quantum operations on our system quickly enough such that cavity decay rate and thermal damping do not effect our system. 
%In other words in the strong coupling regime the non-dissipative part of the Hamiltonian dominates all decay channels $g\gg \kappa, \Gamma_m$ and 
%In the SCR the usual Lorentzian shaped susceptibility $\chi$ of our harmonic oscillator (see Eq.\ref{eq:Sxx}) changes into a double peak  \cite{Dobrindt2008b}, corresponding to two new eigenmodes due to the hybridization of the optical and the mechanical mode. %with the average linewidth  of optical and mechanical linewidth $(\kappa +  \Gamma_m)/2$. Therefore, to resolve the NMS of $2g_y$, $ \Gamma_m$ needs to be smaller or comparable to $\kappa$. 
The hybridized eigenmode frequencies 
\begin{equation}\label{eq:Omega+-}
\Omega_{\pm} =\Omega_{\text{m}} - \frac{\Omega_{\text{m}} + \Delta}{2} \pm \sqrt{g_{\text{y}}^2 + \left(\frac{\Omega_{\text{m}} +\Delta}{2}\right)^2} 
\end{equation}
\noindent experience an avoided crossing, the so-called NMS, which reaches a maximum of $\Omega_+ - \Omega_- = 2 g_\text{y}$ at the optimal detuning  $ \Delta = -\Omega_\text{m}$. The linewidth of the hybrid modes at this detuning is $(\kappa +  \Gamma_\text{m})/2$. % where $\Gamma_m=2\pi\times 0.8$kHz 
Therefore, $ \Gamma_\text{m}$ needs to be smaller or comparable to $\kappa$ to resolve the NMS of $2g_\text{y}$.\\

%%%%%%%%%%%%%%%%%%%%%%%%%%%%%%%%%%%%%%%%%%%%%%%%%%%%%%%%%%%%%%%%%%%%%%%%%%%%%%%%%%%%%%%%%%%%%%%%%%%%%
\begin{figure}[h]
	\begin{center}
		\includegraphics[width=0.5\textwidth]{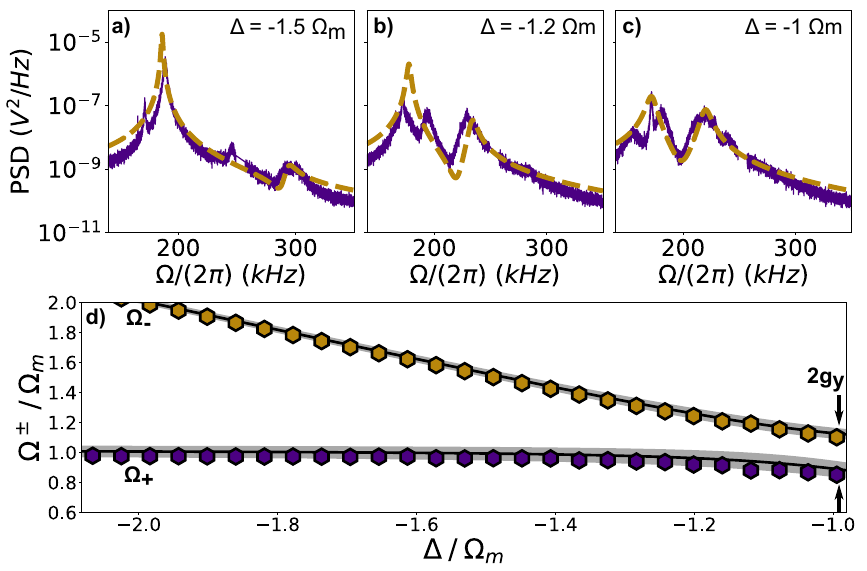}
		\caption{\footnotesize \textbf{Power spectral density versus cavity detuning $\Delta$}  (\textbf{a}-\textbf{c}) Experiment (purple) and theory (yellow, dashed) fitted to Eq.\ref{eq:Sxx} at (\textbf{a})  %$\Delta =  - 2\pi\times 449\text{kHz} \approx -2.3 \Omega_m$, (\textbf{b})
		 $\Delta = -2\pi\times  293\text{kHz} \approx -1.5 \Omega_\text{m}$, (\textbf{b}) $\Delta =-2\pi\times 225\text{kHz} \approx -1.2 \Omega_\text{m}$, and (\textbf{c})$\Delta = -2\pi\times  205\text{kHz} \approx -\Omega_\text{m}$. The optomechanical coupling strength $g_\text{y}$ grows with increasing $\Delta$. Optical and mechanical modes start to hybridise clearly at $\Delta \geq -1.5 \Omega_\text{m}$. % and the peaks split maximally at $\Delta = -2\pi\times 205\text{kHz}\approx -\Omega_m$.
		 We attribute the  discrepancy between data and theory to the second optical mode (see Supplementary Informtation). %Methods \ref{app:MultiModeCoupl}). 
		 (\textbf{d}) Hybridised eigenmodes $\Omega_{\pm}$ versus $\Delta$ at the intensity minimum ($y_0 \approx \lambda_\text{c}/4$). % corresponding to  $\phi = 0.54\pi$.
		 Maximum normal mode splitting of $2g_\text{y}$ with $g_\text{y} =2\pi\times 22.8\text{kHz} = 2.3\kappa$ occurs at $\Delta = -\Omega_\text{m}$. The black line fits the data to Eq.\ref{eq:Omega+-}, while the inner (outer) edges of the grey area correspond to a fit using solely to the upper (lower) branch $\Omega_{-}$ ($\Omega_{+}$).	}
		\label{fig:03}
	\end{center}
\end{figure}
%%%%%%%%%%%%%%%%%%%%%%%%%%%%%%%%%%%%%%%%%%%%%%%%%%%%%%%%%%%%%%%%%%%%%%%%%%%%%%%%%%%%%%%%%%%%%%%%%%%%%

%%%%%%%%%%%%%%%%%%%%%%%%%%%%%%%%%%%%%%%%%%%%%%%%%%%%%%%%%%%%%%%%%%%%%%%%%%%%%%%%%%%%%%%%%%%%%%%%%%%%%
\begin{figure}
	\begin{center}
		\includegraphics[width=0.48\textwidth]{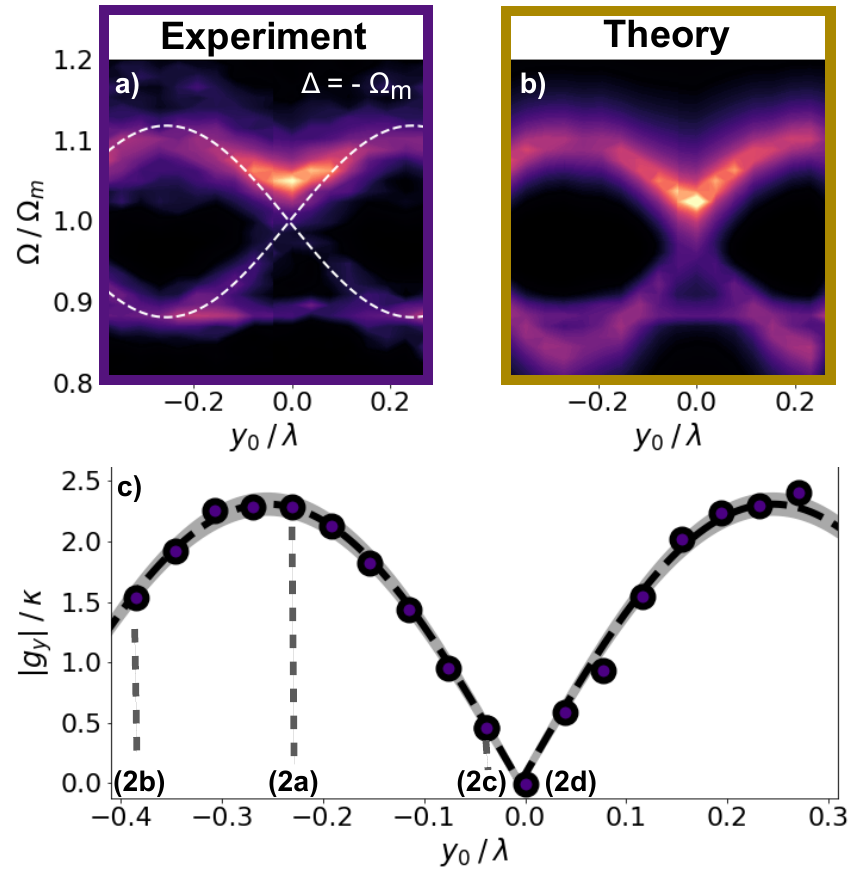}
		\caption{\footnotesize \textbf{Normal Mode Splitting versus particle position $y_0$:} (\textbf{a}) Experiment and (\textbf{b}) theory according to Eq.\ref{eq:Sxx}. Particle position power spectral density PSD$(\Omega)$ at the optimal $\Delta = -2\pi\times 193\text{kHz}  \approx -\Omega_\text{m}$ along $y_0$ is shown. The hybridized modes split by $2g_\text{y}$. The white dashed line displays $\Omega_{\pm}/\Omega_{\text{m}} = 1 \pm g_{\text{y}}/\Omega_{\text{m}}$ where $g_{\text{y}}$ follows Eq.\ref{eq:gy}. The mechanical mode at $\Omega/\Omega_{\text{m}} \approx 0.89$ corresponds to the mechanical $x-$mode. The data and fit show very good agreement. (\textbf{c})  $\vert g_\text{y}\vert$ at $\Delta \approx - \Omega_\text{m}$ versus $y_0$. Maximum and minimum coupling are separated by $\delta y_0 = \lambda_\text{c}/4$ as expected by Eq.\ref{eq:gy}. Black dashed line fits to the absolute value of Eq.\ref{eq:gy} with a maximum $g_{\text{y}}^{\text{max}} \approx -2.3\kappa$ and the grey shaded area corresponds to 3$\sigma_{\text{std}}$ of the fit. The dotted lines indicate the positions used in Fig.\ref{fig:02}.  }
		\label{fig:04}
	\end{center}
\end{figure}
%%%%%%%%%%%%%%%%%%%%%%%%%%%%%%%%%%%%%%%%%%%%%%%%%%%%%%%%%%%%%%%%%%%%%%%%%%%%%%%%%%%%%%%%%%%%%%%%%%%%%

%\paragraph*{\textbf{Results and Discussion}}
\noindent As can be seen from Eq.\ref{eq:gy}, we control $g_{\text{y}}$ through various parameters like the trap power $P_\text{t}$, the particle position $y_0$ and the polarisation angle $\theta$. The optical coupling rate $\Gamma_{\text{opt}}$ depends additionally on the trap detuning $\Delta$ and is maximised at  $\Delta = - \Omega_{\text{m}}$ to  $\Gamma_{\text{opt}}= 4g_{\text{y}}^2/\kappa$ \cite{Aspelmeyer2014,Meyer2019}. While $P_\text{t}$ and $\Delta$ only influence the magnitude of the coupling strength, $y_0$ and $\theta$ change also the nature of the coupling from 1D to potentially 3D \cite{Gonzalez-Ballestero2019}. For simplicity, we focus on varying $\Delta$ and $y_0$ in the following measurements and keep $P_\text{t}$, $\theta$, and $\Gamma_\text{m} = 2\pi\times 0.8$kHz, corresponding to $p=1.4$mBar, fixed (see Methods \ref{app:setup}). 
The range of $\Delta$ is limited due to instabilities in the experiment.  \\

%%%%%%%%%%%%%%%%%%%%%%%%%%%%%%%%%%%%%%%%%%%%%%%%%%%%%%%%%%%%%%%%%%%%%%%%%%%%%%%%%%%%%%%%%%%%%%%%%%%%%

\noindent \textbf{Observation of Strong Coupling}
%\paragraph*{Results}
\noindent Fig.\ref{fig:02} left panel displays the experimental position power spectral density (PSD) versus $\Delta$ for different $y_0$. 
Throughout the remaining part of the manuscript, we fit our PSD to Eq.\ref{eq:Sxx}, if not stated differently. From this fit, we can extract the hybridized modes $\Omega_{\pm}$ that are separated by  $2g_\text{y}$.
%
%, obtained by interfering the trap light with a reference beam.
We cover a total distance of $\delta y_0 %= 287\text{nm}
\approx \lambda_{\text{c}}/4$ and change the optomechanical coupling strength, and therefore also the NMS, from (a) $g_{\text{y}}/2\pi =  22.8\text{kHz}$, (b) $15.4\text{kHz}$, (c) $4.6\text{kHz}$ and (d) $0\text{kHz}$, exploring the entire range from strong coupling to zero coupling. The right panel shows the fit%to Eq.\ref{eq:Sxx}%for all mechanical modes 
, which is in good agreement with the data. We observe  %hybridisation of the mechanical and optical mode to 
two eigenmodes $\Omega_{\pm}$ with an exceptional NMS of $%\Omega_+ -\Omega_- = 
2g_{\text{y}} \approx 4.6 \kappa$ at %the optimal detuning of 
$\Delta = -\Omega_{\text{m}}$, corresponding to 20\% of the bare mechanical eigenfrequency, once the system enters the SCR at $g>\kappa/4 $ \cite{Dobrindt2008b}. %The NMS reduces with decreasing optomechanical coupling $g_y$ until it vanishes completely and
For $g_{\text{y}} =0$, we observe only the mechanical mode with slightly increased frequency $\Omega_{\text{m}} = 2\pi \times 200 \text{kHz}$ due to the additional trapping potential supplied by the cavity field (see Fig.\ref{fig:02}(d)). % at m= 14
%The maximum NMS achieves an exceptional value of $2g_y \approx 4.6 \kappa$ corresponding to 20\% of the bare mechanical eigenfrequency.\\
%The linewidth of the hybrised modes is a weighted mixture of the mechanical damping $\Gamma_m=2\pi\times 0.8$kHz and the cavity linewidth $\kappa$. \\
In Fig.\ref{fig:02}(a) and (b) we observe an additional NMS in the $y$-mode, which stems from a second cross polarised optical mode. Note that, throughout all our measurements (see Fig.\ref{fig:02}-Fig.\ref{fig:04}), the second NMS is the largest source for discrepancies between experiment and theory (for more details see Supplementary Information). % Metohds \ref{app:MultiModeCoupl}). 
We also attribute the NMS of the $x$-mode at $\Omega/\Omega_{\text{m}} = 0.89$  to the second optical mode as observed in Fig.\ref{fig:02}(d), since the $x$-mode should be decoupled from the first mode ($g_{\text{x}} =0$ if $\theta =0$).\\
\noindent Fig.\ref{fig:03}(a)-(c) displays the particle's position PSD at different $\Delta$ while it is located at the intensity minimum, corresponding to the position of maximum coupling $g_{\text{y}}= 2.3\kappa$, displayed in Fig.\ref{fig:02}(a). Our theory (yellow) captures the data (purple) well. %The residual discrepancy between data and theory is mainly due to the second cross polarised optical mode inducing a second NMS. %at $\Delta = -\Omega_m -2\pi\times 34\text{kHz}$. 
In Fig.\ref{fig:03}(a) %the bare mechanical mode is observed at large $\Delta$, while with increasing $\Delta$
the optical mode and mechancial mode begin to hybridise into new eigenmodes at $\Delta = -1.5 \Omega_{\text{m}}$ which is confirmed by a second peak appearing at  $ \Omega \approx 2\pi\times 300$kHz. The hybridization becomes stronger as $\Delta$ approaches the cavity resonance and the NMS is maximized at $\Delta \approx -\Omega_{\text{m}}$ as shown in Fig.\ref{fig:03}(c). The dependence of the new eigenmodes $\Omega_{\pm}$ on $\Delta$ is shown in Fig.\ref{fig:03}(d), displaying clearly the expected avoided crossing of $2g_{\text{y}}$. 	The solid line is a fit to Eq.\ref{eq:Omega+-}. The edges of the shaded area represent the upper and lower limit of the fit, which we obtain by fitting only the upper branch (yellow) or the lower branch (purple), respectively.
\\
\noindent As already discussed previously, our experiment allows to change the optomechanical coupling by changing various experimental parameters, which stands in contrast to many other experimental platforms. Fig.4 displays this flexibility to reach the SCR by demonstrating the position dependence of $g_{\text{y}}$ at optimal detuning %$\Delta = -2\pi\times 193 \text{kHz} \approx -\Omega_m$ 
$\Delta \approx -\Omega_{\text{m}}$ extracted from the data Fig.2(a)-(d).  The experimental and theoretical position PSDs versus $y_0$ are depicted in Fig.\ref{fig:04}(a) and (b).%, where theoretical values are obtained by fitting the data to Eq.\ref{eq:Sxx}. 
The mode at $\Omega/\Omega_{\text{m}} \approx 0.89$ corresponds to the decoupled $x$-mode. The dashed line highlights the theoretical frequency of the eigenmodes $\Omega_{\pm}/\Omega_{\text{m}}$ following Eq.\ref{eq:Omega+-}. In both experiment and theory we observe the expected sinusoidal behaviour predicted by Eq.\ref{eq:gy}. Fig.\ref{fig:04}(c) depicts $\vert g_{\text{y}}\vert =(\Omega_+ -\Omega_-)/2$ (circles) extracted from Fig.\ref{fig:04}(a). The dashed line represents the fit to the absolute value of Eq.\ref{eq:gy} yielding $g_{\text{y}}^{\text{exp}} = 2\pi\times (22.8 \pm 0.2)$kHz which coincides well with the theoretical value of $g_{\text{y}}^{\text{th}}= 2\pi\times 22.6 $kHz. The measured period coincides with the expected period  of $\lambda_{\text{c}}/4$. The shaded area corresponds to 3$\sigma_{\text{std}}$ of the fit.\\

%=================================================================
\subsection*{Discussion}
\noindent As a figure of merit to assess the  potential of our system for quantum applications, we use the quantum cooperativity, which yields here $C_{\text{CS}} = (2 g_{\text{y}}^{\text{max}})^2/(\kappa \Gamma_{\text{m}} (n_{\text{th}}+1)) = 8 \times 10^{-6}$ at a pressure $p = 1.4$mbar and promises a value as large as $C_{\text{CS}} \approx 36$ at $p = 3\times 10^{-7}$mbar, since $\Gamma_{\text{m}} \propto p$. At this low pressure, the photon recoil heating rate $\Gamma_{\text{rec}}$ \cite{Jain2016} equals our mechanical decoherence rate $\Gamma_{\text{m}} (n_{\text{th}}+1))$, and therefore halves the reachable $C_{\text{CS}}$. The maximum $C_{\text{CS}}$ is ultimately limited by $\Gamma_{\text{rec}}$, regardless if we reduce the pressure even further. Nevertheless, this predicted value of $C_{\text{CS}}$ is many orders of magnitude larger than what has been achieved in levitation setups by actively driving the cavity \cite{Meyer2019,Delic2020a} and larger than achieved in \cite{Delic2020}. % and outperforms electromechanical systems by a factor of 10 \cite{Teufel2011a}.
More importantly it enables coherent quantum control at $g \gg \kappa,\Gamma_{\text{m}}\cdot n_{\text{th}}$ at pressure levels $p \leq 10^{-6}$mbar, a pressure regime commonly demonstrated in numerous levitation experiments \cite{Jain2016,Meyer2019,Delic2020}. %\\ At low pressures quantum states can be stored for $\tau = 1/(\Gamma_m \: n_{th}) <100\mu$s before the absorption one thermal phonon from its environment.

Furthermore, our experimental parameters promise the possibility of motional ground state cooling in our system \cite{Delic2020}, which in combination with coherent quantum control enables us to fully enter the quantum regime with levitated systems and to create non-classical states of motion and superposition states of macroscopic objects in free fall experiments \cite{Romero-Isart2011,Bateman2014a} in the future.\\

%%%%%%%%%%%%%%%%%%%%%%%%%%%%%%%%%%%%%%%%%%%%%%%%%%%%%%%%%%%%%%%%%%%%%%%%%%%%%%%%%%%%%%%%%%%%%%%%%%%%%%%%%%

\vspace{0.5cm}

\textbf{Data availability}.\hspace{0.2cm}
The data that support the findings of this study are available from the corresponding author upon request.

\textbf{Acknowledgments}.\hspace{0.2cm} The authors thank J.\:Gieseler and C.\:Gonz\'{a}lez-Ballestero for stimulating discussions. The project acknowledges financial support from the European Research Council through grant QnanoMECA (CoG - 64790), Fundaci\'{o} Privada Cellex, CERCA Programme / Generalitat de Catalunya, and the Spanish Ministry of Economy and Competitiveness through the Severo Ochoa Programme for Centres of Excellence in R$\&$D (SEV-2015-0522). %, grant FIS2016-80293-R. 
This project has received funding from the European Union's Horizon 2020 research and innovation programme under the Marie Sk\l{}odowska-Curie grant agreement No 713729. %LN and VJ acknowledge support through the NCCR-QSIT program (Grant No. 51NF40-160591) by the Swiss National Science Foundation and JG received funding from the European Union's Marie Sk\l{}odowska-Curie program (SEQOO, H2020-MSCA-IF-2014, grant no. 655369).
\\
\textbf{Authors contribution}
A.D.L.R.S. performed the measurements, N.M. evaluated the data and wrote the manuscript, R.Q supervised the study.

\renewcommand{\bibsection}{\section*{References}}
%\bibliographystyle{plainnatnourl}
%\bibliography{bib/ZZ_Publications-NMS}
%\bibliography{X:/Publications/Paper/Bibliographies/ZZ_Publications-NMS}

\begin{thebibliography}{10}
	\expandafter\ifx\csname url\endcsname\relax
	\def\url#1{\texttt{#1}}\fi
	\expandafter\ifx\csname urlprefix\endcsname\relax\def\urlprefix{URL }\fi
	\providecommand{\bibinfo}[2]{#2}
	\providecommand{\eprint}[2][]{\url{#2}}
	
	\bibitem{Chang2010}
	\bibinfo{author}{Chang, D.~E.} \emph{et~al.}
	\newblock \bibinfo{title}{{Cavity opto-mechanics using an optically levitated
			nanosphere}}.
	\newblock \emph{\bibinfo{journal}{PNAS}} \textbf{\bibinfo{volume}{07}},
	\bibinfo{pages}{1005--1010} (\bibinfo{year}{2010}).
	
	\bibitem{Romero-Isart2010}
	\bibinfo{author}{Romero-Isart, O.}, \bibinfo{author}{Juan, M.~L.},
	\bibinfo{author}{Quidant, R.} \& \bibinfo{author}{Cirac, J.~I.}
	\newblock \bibinfo{title}{{Toward quantum superposition of living organisms}}.
	\newblock \emph{\bibinfo{journal}{New Journal of Physics}}
	\textbf{\bibinfo{volume}{12}}, \bibinfo{pages}{033015}
	(\bibinfo{year}{2010}).
	
	\bibitem{Kiesel2013}
	\bibinfo{author}{Kiesel, N.} \emph{et~al.}
	\newblock \bibinfo{title}{{Cavity cooling of an optically levitated submicron
			particle}}.
	\newblock \emph{\bibinfo{journal}{PNAS}} \textbf{\bibinfo{volume}{110}},
	\bibinfo{pages}{35} (\bibinfo{year}{2013}).
	
	\bibitem{Asenbaum2013}
	\bibinfo{author}{Asenbaum, P.}, \bibinfo{author}{Kuhn, S.},
	\bibinfo{author}{Nimmrichter, S.}, \bibinfo{author}{Sezer, U.} \&
	\bibinfo{author}{Arndt, M.}
	\newblock \bibinfo{title}{{Cavity cooling of free silicon nanoparticles in high
			vacuum}}.
	\newblock \emph{\bibinfo{journal}{Nature Communications}}
	\textbf{\bibinfo{volume}{4}}, \bibinfo{pages}{2743} (\bibinfo{year}{2013}).
	
	\bibitem{Millen2015}
	\bibinfo{author}{Millen, J.}, \bibinfo{author}{Fonseca, P. Z.~G.},
	\bibinfo{author}{Mavrogordatos, T.}, \bibinfo{author}{Monteiro, T.~S.} \&
	\bibinfo{author}{Barker, P.~F.}
	\newblock \bibinfo{title}{{Cavity Cooling a Single Charged Levitated
			Nanosphere}}.
	\newblock \emph{\bibinfo{journal}{Physical Review Letters}}
	\textbf{\bibinfo{volume}{114}}, \bibinfo{pages}{123602}
	(\bibinfo{year}{2015}).
	
	\bibitem{Fonseca2016}
	\bibinfo{author}{Fonseca, P. Z.~G.}, \bibinfo{author}{Aranas, E.~B.},
	\bibinfo{author}{Millen, J.}, \bibinfo{author}{Monteiro, T.~S.} \&
	\bibinfo{author}{Barker, P.~F.}
	\newblock \bibinfo{title}{{Nonlinear Dynamics and Strong Cavity Cooling of
			Levitated Nanoparticles}}.
	\newblock \emph{\bibinfo{journal}{Physical Review Letters}}
	\textbf{\bibinfo{volume}{117}}, \bibinfo{pages}{173602}
	(\bibinfo{year}{2016}).
	
	\bibitem{Meyer2019}
	\bibinfo{author}{Meyer, N.} \emph{et~al.}
	\newblock \bibinfo{title}{{Resolved-Sideband Cooling of a Levitated
			Nanoparticle in the Presence of Laser Phase Noise}}.
	\newblock \emph{\bibinfo{journal}{Physical Review Letters}}
	\textbf{\bibinfo{volume}{123}}, \bibinfo{pages}{1--6} (\bibinfo{year}{2019}).
	
	\bibitem{Delic2019}
	\bibinfo{author}{Deli{\'{c}}, U.} \emph{et~al.}
	\newblock \bibinfo{title}{{Cavity Cooling of a Levitated Nanosphere by Coherent
			Scattering}}.
	\newblock \emph{\bibinfo{journal}{Physical Review Letters}}
	\textbf{\bibinfo{volume}{122}}, \bibinfo{pages}{123602}
	(\bibinfo{year}{2019}).
	
	\bibitem{Delic2020}
	\bibinfo{author}{Deli{\'{c}}, U.} \emph{et~al.}
	\newblock \bibinfo{title}{{Cooling of a levitated nanoparticle to the motional
			quantum ground state}}.
	\newblock \emph{\bibinfo{journal}{Science}} \textbf{\bibinfo{volume}{367}},
	\bibinfo{pages}{892--895} (\bibinfo{year}{2020}).
	
	\bibitem{Romero-Isart2011}
	\bibinfo{author}{Romero-Isart, O.} \emph{et~al.}
	\newblock \bibinfo{title}{{Large Quantum Superpositions and Interference of
			Massive Nanometer-Sized Objects}}.
	\newblock \emph{\bibinfo{journal}{Physical Review Letters}}
	\textbf{\bibinfo{volume}{107}}, \bibinfo{pages}{020405}
	(\bibinfo{year}{2011}).
	
	\bibitem{Bateman2014a}
	\bibinfo{author}{Bateman, J.}, \bibinfo{author}{Nimmrichter, S.},
	\bibinfo{author}{Hornberger, K.} \& \bibinfo{author}{Ulbricht, H.}
	\newblock \bibinfo{title}{{Near-field interferometry of a free-falling
			nanoparticle from a point-like source}}.
	\newblock \emph{\bibinfo{journal}{Nature Communications}}
	\textbf{\bibinfo{volume}{5}}, \bibinfo{pages}{1--5} (\bibinfo{year}{2014}).
	
	\bibitem{Aspelmeyer2014}
	\bibinfo{author}{Aspelmeyer, M.}, \bibinfo{author}{Kippenberg, T.~J.} \&
	\bibinfo{author}{Marquardt, F.}
	\newblock \bibinfo{title}{{Cavity optomechanics}}.
	\newblock \emph{\bibinfo{journal}{Reviews of Modern Physics}}
	\textbf{\bibinfo{volume}{86}}, \bibinfo{pages}{1391--1452}
	(\bibinfo{year}{2014}).
	
	\bibitem{Millen2020}
	\bibinfo{author}{Millen, J.}, \bibinfo{author}{Monteiro, T.~S.},
	\bibinfo{author}{Pettit, R.} \& \bibinfo{author}{Vamivakas, A.~N.}
	\newblock \bibinfo{title}{{Optomechanics with levitated particles}}.
	\newblock \emph{\bibinfo{journal}{Reports on Progress in Physics}}
	\textbf{\bibinfo{volume}{83}} (\bibinfo{year}{2020}).
	
	\bibitem{Hempston2017}
	\bibinfo{author}{Hempston, D.} \emph{et~al.}
	\newblock \bibinfo{title}{{Force sensing with an optically levitated charged
			nanoparticle}}.
	\newblock \emph{\bibinfo{journal}{Applied Physics Letters}}
	\textbf{\bibinfo{volume}{111}}, \bibinfo{pages}{133111}
	(\bibinfo{year}{2017}).
	
	\bibitem{Arita2013}
	\bibinfo{author}{Arita, Y.}, \bibinfo{author}{Mazilu, M.} \&
	\bibinfo{author}{Dholakia, K.}
	\newblock \bibinfo{title}{{Laser-induced rotation and cooling of a trapped
			microgyroscope in vacuum}}.
	\newblock \emph{\bibinfo{journal}{Nature Communications}}
	\textbf{\bibinfo{volume}{4}}, \bibinfo{pages}{2374} (\bibinfo{year}{2013}).
	
	\bibitem{Kuhn2017}
	\bibinfo{author}{Kuhn, S.} \emph{et~al.}
	\newblock \bibinfo{title}{{Optically driven ultra-stable nanomechanical
			rotor}}.
	\newblock \emph{\bibinfo{journal}{Nature Communications}}
	\textbf{\bibinfo{volume}{8}}, \bibinfo{pages}{1670} (\bibinfo{year}{2017}).
	
	\bibitem{Monteiro2018}
	\bibinfo{author}{Monteiro, F.}, \bibinfo{author}{Ghosh, S.},
	\bibinfo{author}{van Assendelft, E.~C.} \& \bibinfo{author}{Moore, D.~C.}
	\newblock \bibinfo{title}{{Optical rotation of levitated spheres in high
			vacuum}}.
	\newblock \emph{\bibinfo{journal}{Physical Review A}}
	\textbf{\bibinfo{volume}{97}}, \bibinfo{pages}{051802(R)}
	(\bibinfo{year}{2018}).
	
	\bibitem{Reimann2018}
	\bibinfo{author}{Reimann, R.} \emph{et~al.}
	\newblock \bibinfo{title}{{GHz Rotation of an Optically Trapped Nanoparticle in
			Vacuum}}.
	\newblock \emph{\bibinfo{journal}{Physical Review Letters}}
	\textbf{\bibinfo{volume}{121}}, \bibinfo{pages}{033602}
	(\bibinfo{year}{2018}).
	
	\bibitem{Hebestreit2018a}
	\bibinfo{author}{Hebestreit, E.}, \bibinfo{author}{Frimmer, M.},
	\bibinfo{author}{Reimann, R.} \& \bibinfo{author}{Novotny, L.}
	\newblock \bibinfo{title}{{Sensing Static Forces with Free-Falling
			Nanoparticles}}.
	\newblock \emph{\bibinfo{journal}{Physical Review Letters}}
	\textbf{\bibinfo{volume}{121}}, \bibinfo{pages}{063602}
	(\bibinfo{year}{2018}).
	
	\bibitem{Rondin2017}
	\bibinfo{author}{Rondin, L.} \emph{et~al.}
	\newblock \bibinfo{title}{{Direct measurement of Kramers turnover with a
			levitated nanoparticle}}.
	\newblock \emph{\bibinfo{journal}{Nature Nanotechnology}}
	\textbf{\bibinfo{volume}{12}}, \bibinfo{pages}{1130--1134}
	(\bibinfo{year}{2017}).
	
	\bibitem{Marshall2003}
	\bibinfo{author}{Marshall, W.}, \bibinfo{author}{Simon, C.},
	\bibinfo{author}{Penrose, R.} \& \bibinfo{author}{Bouwmeester, D.}
	\newblock \bibinfo{title}{{Towards Quantum Superpositions of a Mirror}}.
	\newblock \emph{\bibinfo{journal}{Physcal Review Letters}}
	\textbf{\bibinfo{volume}{91}}, \bibinfo{pages}{13} (\bibinfo{year}{2003}).
	
	\bibitem{Kleckner2008}
	\bibinfo{author}{Kleckner, D.} \emph{et~al.}
	\newblock \bibinfo{title}{{Creating and verifying a quantum superposition in a
			micro-optomechanical system}}.
	\newblock \emph{\bibinfo{journal}{New Journal of Physics}}
	\textbf{\bibinfo{volume}{10}}, \bibinfo{pages}{095020}
	(\bibinfo{year}{2008}).
	
	\bibitem{Windey2019}
	\bibinfo{author}{Windey, D.} \emph{et~al.}
	\newblock \bibinfo{title}{{Cavity-Based 3D Cooling of a Levitated Nanoparticle
			via Coherent Scattering}}.
	\newblock \emph{\bibinfo{journal}{Physical Review Letters}}
	\textbf{\bibinfo{volume}{122}}, \bibinfo{pages}{123601}
	(\bibinfo{year}{2019}).
	
	\bibitem{Delic2020a}
	\bibinfo{author}{Deli{\'{c}}, U.} \emph{et~al.}
	\newblock \bibinfo{title}{{Levitated cavity optomechanics in high vacuum}}.
	\newblock \emph{\bibinfo{journal}{Quantum Science and Technology}}
	\textbf{\bibinfo{volume}{5}}, \bibinfo{pages}{025006} (\bibinfo{year}{2020}).
	
	\bibitem{Safavi2013}
	\bibinfo{author}{Safavi-Naeini, A.~H.} \emph{et~al.}
	\newblock \bibinfo{title}{{Squeezed light from a silicon micromechanical
			resonator}}.
	\newblock \emph{\bibinfo{journal}{Nature Letter}}
	\textbf{\bibinfo{volume}{500}}, \bibinfo{pages}{185--189}
	(\bibinfo{year}{2013}).
	
	\bibitem{Riedinger2016}
	\bibinfo{author}{Riedinger, R.} \emph{et~al.}
	\newblock \bibinfo{title}{{Non-classical correlations between single photons
			and phonons from a mechanical oscillator}}.
	\newblock \emph{\bibinfo{journal}{Nature}} \textbf{\bibinfo{volume}{530}},
	\bibinfo{pages}{313--316} (\bibinfo{year}{2016}).
	
	\bibitem{Riedinger2018}
	\bibinfo{author}{Riedinger, R.} \emph{et~al.}
	\newblock \bibinfo{title}{{Remote quantum entanglement between two
			micromechanical oscillators}}.
	\newblock \emph{\bibinfo{journal}{Nature}} \textbf{\bibinfo{volume}{556}},
	\bibinfo{pages}{473--477} (\bibinfo{year}{2018}).
	
	\bibitem{Chen2020}
	\bibinfo{author}{Chen, J.}, \bibinfo{author}{Rossi, M.},
	\bibinfo{author}{Mason, D.} \& \bibinfo{author}{Schliesser, A.}
	\newblock \bibinfo{title}{{Entanglement of propagating optical modes via a
			mechanical interface}}.
	\newblock \emph{\bibinfo{journal}{Nature Communications}}
	\textbf{\bibinfo{volume}{11}} (\bibinfo{year}{2020}).
	
	\bibitem{Groblacher2009}
	\bibinfo{author}{Gr{\"{o}}blacher, S.}, \bibinfo{author}{Hammerer, K.},
	\bibinfo{author}{Vanner, M.~R.} \& \bibinfo{author}{Aspelmeyer, M.}
	\newblock \bibinfo{title}{{Observation of strong coupling between a
			micromechanical resonator and an optical cavity field}}.
	\newblock \emph{\bibinfo{journal}{Nature}} \textbf{\bibinfo{volume}{460}},
	\bibinfo{pages}{724--727} (\bibinfo{year}{2009}).
	
	\bibitem{Teufel2011a}
	\bibinfo{author}{Teufel, J.~D.} \emph{et~al.}
	\newblock \bibinfo{title}{{Circuit cavity electromechanics in the
			strong-coupling regime}}.
	\newblock \emph{\bibinfo{journal}{Nature}} \textbf{\bibinfo{volume}{471}},
	\bibinfo{pages}{204--208} (\bibinfo{year}{2011}).
	
	\bibitem{Teufel2011}
	\bibinfo{author}{Teufel, J.~D.} \emph{et~al.}
	\newblock \bibinfo{title}{{Sideband cooling of micromechanical motion to the
			quantum ground state}}.
	\newblock \emph{\bibinfo{journal}{Nature Letter}}
	\textbf{\bibinfo{volume}{475}}, \bibinfo{pages}{359--363}
	(\bibinfo{year}{2011}).
	
	\bibitem{Verhagen2012c}
	\bibinfo{author}{Verhagen, E.}, \bibinfo{author}{Del{\'{e}}glise, S.},
	\bibinfo{author}{Weis, S.}, \bibinfo{author}{Schliesser, A.} \&
	\bibinfo{author}{Kippenberg, T.~J.}
	\newblock \bibinfo{title}{{Quantum-coherent coupling of a mechanical oscillator
			to an optical cavity mode}}.
	\newblock \emph{\bibinfo{journal}{Nature}} \textbf{\bibinfo{volume}{482}},
	\bibinfo{pages}{63--67} (\bibinfo{year}{2012}).
	
	\bibitem{Dobrindt2008b}
	\bibinfo{author}{Dobrindt, J.~M.}, \bibinfo{author}{Wilson-Rae, I.} \&
	\bibinfo{author}{Kippenberg, T.~J.}
	\newblock \bibinfo{title}{{Parametric normal-mode splitting in cavity
			optomechanics}}.
	\newblock \emph{\bibinfo{journal}{Physical Review Letters}}
	\textbf{\bibinfo{volume}{101}}, \bibinfo{pages}{1--4} (\bibinfo{year}{2008}).
	
	\bibitem{Thompson1992}
	\bibinfo{author}{Thompson, R.~J.}, \bibinfo{author}{Rempe, G.} \&
	\bibinfo{author}{Kimble, H.~J.}
	\newblock \bibinfo{title}{{Observation of Normal-Mode Splitting for an Atom in
			an Optical Cavity}}.
	\newblock \emph{\bibinfo{journal}{Physical Review Letters}}
	\textbf{\bibinfo{volume}{68}}, \bibinfo{pages}{1132--1135}
	(\bibinfo{year}{1992}).
	
	\bibitem{Vuletic2001}
	\bibinfo{author}{Vuleti{\'{c}}, V.}, \bibinfo{author}{Chan, H.~W.} \&
	\bibinfo{author}{Black, A.~T.}
	\newblock \bibinfo{title}{{Three-dimensional cavity Doppler cooling and cavity
			sideband cooling by coherent scattering}}.
	\newblock \emph{\bibinfo{journal}{Physcial Review A}}
	\textbf{\bibinfo{volume}{64}}, \bibinfo{pages}{033405}
	(\bibinfo{year}{2001}).
	
	\bibitem{Gonzalez-Ballestero2019}
	\bibinfo{author}{Gonzalez-Ballestero, C.} \emph{et~al.}
	\newblock \bibinfo{title}{{Theory for Cavity Cooling of Levitated Nanoparticles
			via Coherent Scattering: Master Equation Approach}}.
	\newblock \emph{\bibinfo{journal}{Physical Review A}}
	\textbf{\bibinfo{volume}{100}}, \bibinfo{pages}{013805}
	(\bibinfo{year}{2019}).
	
	\bibitem{Chikkaraddy2016}
	\bibinfo{author}{Chikkaraddy, R.} \emph{et~al.}
	\newblock \bibinfo{title}{{Single-molecule strong coupling at room temperature
			in plasmonic nanocavities}}.
	\newblock \emph{\bibinfo{journal}{Nature}} \textbf{\bibinfo{volume}{535}},
	\bibinfo{pages}{127--130} (\bibinfo{year}{2016}).
	
	\bibitem{Kleemann2017a}
	\bibinfo{author}{Kleemann, M.~E.} \emph{et~al.}
	\newblock \bibinfo{title}{{Strong-coupling of WSe2 in ultra-compact plasmonic
			nanocavities at room temperature}}.
	\newblock \emph{\bibinfo{journal}{Nature Communications}}
	\textbf{\bibinfo{volume}{8}} (\bibinfo{year}{2017}).
	
	\bibitem{Gieseler2014a}
	\bibinfo{author}{Gieseler, J.}
	\newblock \emph{\bibinfo{title}{{Dynamics of optically levitated nanoparticles
				in high vacuum}}}.
	\newblock Ph.D. thesis (\bibinfo{year}{2014}).
	\newblock
	%\urlprefix\url{http://upcommons.upc.edu//handle/10803/144555{\%}5Cnhttp://upcommons.upc.edu/handle/10803/144555}.
	
	\bibitem{Tebbenjohanns2020}
	\bibinfo{author}{Tebbenjohanns, F.}, \bibinfo{author}{Frimmer, M.},
	\bibinfo{author}{Jain, V.}, \bibinfo{author}{Windey, D.} \&
	\bibinfo{author}{Novotny, L.}
	\newblock \bibinfo{title}{{Motional Sideband Asymmetry of a Nanoparticle
			Optically Levitated in Free Space}}.
	\newblock \emph{\bibinfo{journal}{Physical Review Letters}}
	\textbf{\bibinfo{volume}{124}}, \bibinfo{pages}{13603}
	(\bibinfo{year}{2020}).
	
	\bibitem{Jain2016}
	\bibinfo{author}{Jain, V.} \emph{et~al.}
	\newblock \bibinfo{title}{{Direct Measurement of Photon Recoil from a Levitated
			Nanoparticle}}.
	\newblock \emph{\bibinfo{journal}{Physical Review Letters}}
	\textbf{\bibinfo{volume}{116}}, \bibinfo{pages}{243601}
	(\bibinfo{year}{2016}).
	
	\bibitem{Mestres2015}
	\bibinfo{author}{Mestres, P.} \emph{et~al.}
	\newblock \bibinfo{title}{{Cooling and manipulation of a levitated nanoparticle
			with an optical fiber trap}}.
	\newblock \emph{\bibinfo{journal}{Applied Physics Letters}}
	\textbf{\bibinfo{volume}{107}}, \bibinfo{pages}{151102}
	(\bibinfo{year}{2015}).
	
\end{thebibliography}

\clearpage
\newpage
%\appendix
\section*{Methods}
\subsection{Experimental Setup}\label{app:setup}
\begin{figure}[h]
	\begin{center}
		\includegraphics[width=0.45\textwidth]{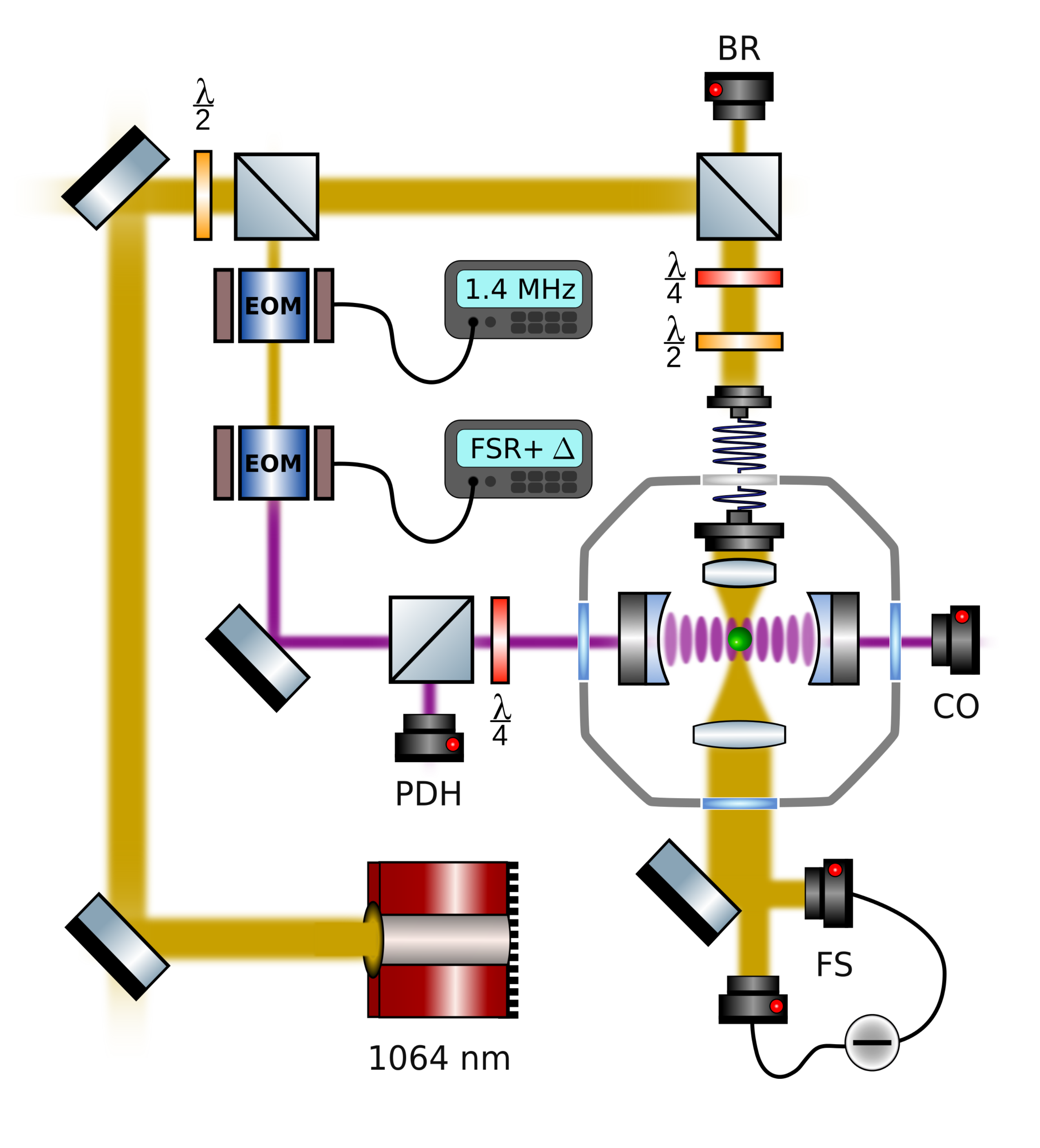}
		\caption{\footnotesize \textbf{Extended experimental setup} (\textbf{a}) A 1064nm Mephisto laser (yellow) traps a silica particle of $d= 177$nm inside a high finesse cavity (purple). The trap light is locked at a variable detuning $\Delta +\text{FSR}$ from the cavity resonance via the Pound-Drever-Hall technique by detecting the error signal on a photodiode (PDH). The particle motion is detected in backreflection (BR) and balanced forward detection (FS). The intracavity field is estimated from the transmitted power detected on a photodiode (CO).}
		\label{fig:app_setup}
	\end{center}
\end{figure}

\noindent The experimental setup is displayed in Fig.\ref{fig:app_setup}. A silica nanoparticle is loaded at ambient pressure into a long range single beam trap and transferred to a more stable, short range optical tweezers trap \cite{Mestres2015} (with wavelength $\lambda_{\text{t}} = 1064~{\rm nm}$, power $P\simeq 150~{\rm mW}$, focusing lens $\text{NA}=0.8$) inside a vacuum chamber. Due to the tight focusing the nanoparticle non-degenerate eigenfrequencies are $\Omega_{\text{x,y,z}} = 2\pi\times ({\rm 172kHz, 197kHz, 56kHz})$ respectively. The optical tweezers are mounted on a 3D nanometer resolution piezo system allowing for precise 3D positioning inside a high finesse Fabry-P\'{e}rot cavity (with cavity finesse $F = 540000$, free spectral range FSR = $2\pi\times 5.4$GHz).\\

\noindent In order to control the detuning $\Delta =  \omega_{\text{t}} - \omega_{\text{c}}$ between the cavity resonance $\omega_{\text{c}}$ and the trap field $\omega_\text{t}$, we use a weak cavity field for locking the cavity via the Pound-Drever-Hall technique (PDH) on the $\text{TEM}_{01}$ mode minimising additional heating effects through the photon recoil heating of the cavity lock field. The PDH errorsignal acts on the internal laser piezo and an external AOM (not shown). We separate lock and trap light in frequency space by one free spectral range (FSR) such that the total detuning between lock and trap yields $\omega_{\text{t}} = \omega - \text{FSR} - \Delta $. The variable EOM modulation $ FSR + \Delta$ is provided by a signal generator. The intracavity power can be deduced from the transmitted cooling light observed on a photodiode behind the cavity (CO). \\
\noindent 
% detection
All particle information shown is gained in forward balanced detection interfering the scattered light field and the non-interacting part of the trap beam as shown in Fig.\ref{fig:app_setup}. The highly divergent trap light is collected using a lens ($\text{NA}=0.8$). We use three balanced detectors (FS) to monitor the oscillation of the particle in all three degrees of freedom. 
\\
The data time traces are acquired at 1MHz acquisition rate. Each particle position PSD is obtained by averaging over N = 25 samples of which each one is calculated from individual 40ms time traces, corresponding to a total measurement time of $t=1$s. 
\\
\noindent We keep the pressure stable at $\text{p} = 1.4$mbar. The thermal bath couples as 
\begin{equation}\label{eq_Gamma_m}
\Gamma_{\text{m}}  =  \frac{k_B T }{\hbar Q_{\text{m}}  n_{\text{th}}}  = 15.8 \frac{r^2 p}{m v_{\text{gas}}}\\
\end{equation}

where $Q_{\text{m}}= \Omega_{\text{m}}/\Gamma_{\text{m}}$ is the mechanical quality factor, $n_{\text{th}} = \frac{k_B T}{\hbar \Omega_{\text{m}}}$ the thermal occupation number, $r$ the particle radius, $p$ the surrounding gas pressure and $v_{\text{gas}} = \sqrt{3 k_B T/m_{\text{gas}}}$.\\
In the measurements presented, we cool our particle's center of mass motion to $T=235$, corresponding to a reduction of the phonon occupation by roughly 20\%. The theoretically expected heating rate due to the residual gas accounts fully for the experimentally observed heating rate.

\subsection{Interaction Hamiltonian and Power Spectral Densities}\label{app:EqMotionPSD}
\noindent Following \cite{Gonzalez-Ballestero2019}, the relevant contributions to the CS interaction Hamiltonian for $\theta = 0$ are given by 

\begin{eqnarray}
\frac{\hat{\rm H}_{\text{CS}}}{\hbar}&=& - g_{\text{y}} (\hat{a}^\dagger +\hat{a})  (\hat{b_y}^\dagger + \hat{b_y})  \nonumber \\& & - g_{\text{z}} (\hat{a}^\dagger - \hat{a}) (\hat{b_z}^\dagger + \hat{b_z})   \\  
\frac{\hat{\rm H}_{\text{DR}}}{\hbar}&=& - g^\text{dr}_{\text{y}}\:\hat{a}^\dagger \hat{a}\:  (\hat{b_y}^\dagger + \hat{b_y})   \\
\frac{\hat{\rm H}_{\text{CAV}}}{\hbar} &=& -  \frac{G_{\perp}}{2} (\:\hat{a}^\dagger + \hat{a}) \cos{\phi}
\end{eqnarray}

\noindent The single photon optomechanical coupling strength achieved by actively driving the cavity is $g^\text{dr}_{\text{y}} = \frac{\alpha\omega_{\text{c}}}{2\epsilon_0 V_\text{c}} \:k_{\text{c}} y_{\text{zpf}} \:\sin{(2\phi)} = 2\pi\times 0.05\text{Hz} \:\sin{(2\phi)}$. This value is enhanced by the intracavity photon number $n_{\text{cav}} =  1.6\times 10^8$, inferred from the transmitted cavity power $P_{\text{out}}$. At optimal conditions, we achieve $g^\text{rp}_{\text{y}} \sqrt{n_{\text{c}}} = 2\pi\times 0.6$kHz. 
Thus, the optomechanical coupling strength is about 40 times larger for CS than for RP, since the photons contributing to the CS interaction are confined in a much smaller volume due to the much smaller trap waist $\text{w}_t \cdot \text{w}_c \ll \text{w}_c^2$.\\ 

\noindent The mechanical susceptibility $\chi$  is given as \cite{Dobrindt2008b}

\begin{eqnarray}\label{eq:Sxx}
	\vert\chi(\Omega)\vert^2 =
& \frac{1}{m^2 [(\Omega_{\text{m}}^2 + 2 \Omega\: \delta\Omega_{\text{m}}(\Omega) - \Omega^2)^2 + (\Omega \Gamma_{\text{eff}}(\Omega))^2]}\\ 
%\chi^{-1}  (\Omega) &=& m [\Omega_m^2 + 2 \Omega\: \delta\Omega_m(\Omega) - \Omega^2 - i\Omega \Gamma_{\text{eff}}(\Omega)]\\
\Gamma_{\text{eff}}(\Omega) = &\Gamma_{\text{m}} + \Gamma_{\text{opt}}(\Omega) \\
\delta\Omega_{\text{m}}(\Omega) =& g_{\text{y}}^2 \frac{\Omega_{\text{m}}}{\Omega} \left[\frac{\Delta + \Omega}{(\Delta + \Omega)^2+ \kappa^2/4} + \frac{\Delta - \Omega}{(\Delta - \Omega)^2+ \kappa^2/4}\right]\\
\Gamma_{\text{opt}}(\Omega) =& g_{\text{y}}^2 \frac{\Omega_{\text{m}}}{\Omega} \left[\frac{\kappa}{(\Delta + \Omega)^2+ \kappa^2/4} - \frac{\kappa}{(\Delta - \Omega)^2+ \kappa^2/4}\right] 
\end{eqnarray}

with the effective (optical) damping $\Gamma_{\text{eff}}$ ($\Gamma_{\text{opt}}$) and the optomechanical spring effect $\delta\Omega_{\text{m}}$. We fit the three mechanical modes $\Omega_{\text{x,y,z}}$ to Eq.\ref{eq:Sxx} where  $g_{\text{y}}$, $\kappa$, $\Gamma_{\text{m}}$ %(initial guess fits very well), 
and the relative mode amplitudes are chosen as free fit parameters. %rest $\pm 5\%$

%%%%%%%%%%%%%%%%%%%%%%%%%%%%%%%%%%%
%%%%%%%%%%%%%%%%%%%%%%%%%%%%%%%%%%%
%%%%%%%%%%%%%%%%%%%%%%%%%%%%%%%%%%%
%%%%%%%%%%%%%%%%%%%%%%%%%%%%%%%%%%%
%%%%%%%%%%%%%%%%%%%%%%%%%%%%%%%%%%%
%%%%%%%%%%%%%%%%%%%%%%%%%%%%%%%%%%%
%\nocite{Toros2020,Dong2012,Sheng2020,Riedinger2018,Chen2020}

\end{document}